\documentclass[12pt]{article}
\pdfoutput=1
\usepackage[T1]{fontenc}
\usepackage{graphicx,psfrag,epsf,color}
\usepackage{amsmath,amssymb,amsfonts}
\usepackage{physics}
\usepackage{siunitx}
\usepackage{array}
\usepackage{cite}
\usepackage{rotating}
\usepackage{slashed, cancel}
\usepackage{xcolor}
\usepackage{subcaption}
\usepackage{hyperref}
\usepackage{graphicx}
\usepackage{mathrsfs}
\usepackage{float}
\restylefloat{table}

\newcommand{\adg}{\mathcal{A}_{\Delta\Gamma_s}^{\mu\mu}}
\newcommand{\smumu}{\mathcal{S}_{\mu\mu}}
\newcommand{\cmumu}{\mathcal{C}_{\mu\mu}}
\newcommand{\cplus}{\mathcal{C}^+}
\newcommand{\cminus}{\mathcal{C}^-}
\usepackage{tabularx}
\usepackage{booktabs}
\usepackage[margin=1.0in]{geometry}
\newcolumntype{C}{>{\centering\arraybackslash}X}
\usepackage{multirow} \usepackage{array}
\title{}
\author{andersrehult}
\date{}

\DeclareUnicodeCharacter{2212}{\ensuremath{-}}

\begin{document}

\begin{titlepage}

\vspace*{-2.0truecm}

\begin{flushright}
Nikhef-2024-008\\
SI-HEP-2024-10\\
P3H-24-027  
\end{flushright}

\vspace*{1.3truecm}

\begin{center}
{
\Large \bf \boldmath Targeting (Pseudo)-Scalar CP Violation with $B_s \to \mu^+\mu^-$}
\end{center}
\vspace{0.9truecm}

\begin{center}
{\bf Robert Fleischer\,${}^{a,b}$,  Eleftheria Malami\,${}^{a,c}$, Anders Rehult\,${}^{a,b}$, and  K. Keri Vos\,${}^{a,d}$}

\vspace{0.5truecm}

${}^a${\sl Nikhef, Science Park 105, NL-1098 XG Amsterdam,  Netherlands}

${}^b${\sl  Department of Physics and Astronomy, Vrije Universiteit Amsterdam,\\
NL-1081 HV Amsterdam, Netherlands}

${}^c${\sl  Center for Particle Physics Siegen (CPPS), Theoretische Physik 1,\\
Universität Siegen, D-57068 Siegen, Germany}

{\sl $^d$Gravitational 
Waves and Fundamental Physics (GWFP),\\ 
Maastricht University, Duboisdomein 30,\\ 
NL-6229 GT Maastricht, the
Netherlands}\\[0.3cm]

\end{center}

\vspace*{1.7cm}

\begin{abstract}
\noindent
The leptonic decay $B_s \to \mu^+\mu^-$ is both rare and theoretically clean, making it an excellent probe for New Physics searches. Due to its helicity suppression in the Standard Model, this decay is particularly sensitive to new (pseudo)-scalar contributions. We present a new strategy for detecting CP-violating New Physics contributions of this kind, exploiting two observables: $\mathcal{A}_{\Delta\Gamma_s}^{\mu\mu}$, which is accessible due to the sizeable decay width difference of the $B_s$ system, and the mixing-induced CP asymmetry $\mathcal{S}_{\mu\mu}$. The strategy also uses information from $B\to K^{(*)} \mu^+\mu^-$ and $B_s\to \phi \mu^+\mu^-$ decays. We find remarkably constrained regions in the $\mathcal{A}_{\Delta\Gamma_s}^{\mu\mu}$--$\mathcal{S}_{\mu\mu}$ plane that serve as promising targets for future measurements.
\end{abstract}


\vspace*{2.1truecm}

\vfill

\noindent
May 2024

\end{titlepage}


\clearpage

\thispagestyle{empty}

\vbox{}

\setcounter{page}{0}

\newpage

\section{Introduction}

The leptonic decay $B_s \to \mu^+\mu^-$ is exceptionally rare -- out of every billion $B_s$ mesons produced, only about three decay into the $\mu^+\mu^-$ final state \cite{ParticleDataGroup:2022pth}. This rarity has made the decay an elusive target. First constraints on its branching ratio were obtained at the Tevatron by the CDF \cite{CDF:2013ezj} and D{{\O}} \cite{D0:2013vkd} collaborations, followed by an observation at the LHC, in a joint effort by LHCb and CMS \cite{CMS:2014xfa}. Recent measurements by LHCb \cite{LHCb:2021awg}, CMS \cite{CMS:2022mgd}, and ATLAS \cite{ATLAS:2018cur} have pushed the experimental precision on the branching ratio to the ten-percent level, a precision set to improve even further in the coming decade \cite{LHCb:2018roe}. Pioneering analyses have also been performed by LHCb \cite{LHCb:2021awg}, CMS \cite{CMS:2022dbz}, and ATLAS \cite{ATLAS:2023trk} on the $B^0_s \to \mu^+\mu^-$ effective lifetime $\tau_s^{\mu\mu}$, measurable due to the sizeable decay width difference $\Delta \Gamma_s$ of the $B_s$ system \cite{DeBruyn:2012wk}. The effective lifetime can be converted into the observable $\adg$.

In the Standard Model (SM), the decay is a flavor-changing neutral current process and is helicity suppressed: its branching ratio is proportional to the muon mass squared. Possible new axial-vector, scalar and pseudo-scalar leptonic currents would affect this decay, where the (pseudo)-scalar interactions would lift the helicity suppression. The rarity of the $B_s \to \mu^+\mu^-$ decay thus makes it an excellent choice for indirect New Physics (NP) searches.  

In these explorations, both the branching ratio and $\adg$ are important observables, and further data on them are highly valuable. However, $B_s \to \mu^+\mu^-$ has even more information to offer. Theoretical work has identified the mixing-induced CP asymmetry $\smumu$ as a promising target of study \cite{Fleischer:2017yox,Fleischer:2017ltw}. This observable, accessible through a time-dependent, flavour-tagged analysis, is generated through interference between $\bar B^0_s \to \mu^+\mu^-$ and $B^0_s \to \mu^+\mu^-$ decay processes that originates from $B^0_s$--$\bar B^0_s$ mixing. However, as pointed out in \cite{Fleischer:2017yox}, the $B^0_s \to \mu^+\mu^-$ observables do not provide sufficient information to determine the relevant NP coefficients. It is only possible to extract them as a function of an unknown parameter.

In this paper, we explore new ways to probe possible (pseudo)-scalar contributions by fully exploiting the rare $B_s\to \mu^+\mu^-$ decay. To accomplish this, we need input on possible NP axial-vector interactions. If present, such contributions as well as new (pseudo)-scalar terms would also enter semileptonic rare $b\to s$ transitions. We demonstrate that (pseudo)-scalar interactions are very suppressed in the semileptonic rare $B$-meson decays. Therefore, these decays can be used to determine the axial-vector NP contributions, letting us focus the analysis of $B^0_s \to \mu^+\mu^-$ on (pseudo)-scalar effects.

We present a new strategy to determine whether such possible (pseudo)-scalar NP terms violate CP symmetry. This method exploits the complementary dependence of $\adg$ and $\smumu$ on such NP effects. In the case of CP-conserving (pseudo)-scalar NP, these observables are strongly correlated: we identify remarkably constrained regions in the $\adg$--$\smumu$ plane that serve as interesting target regions. Future measurements of these observables outside this target region would indicate CP-violating phases in the (pseudo)-scalar interactions. 

This paper is structured as follows: In Sec.~\ref{ch:theoretical_framework}, we introduce our theoretical framework for $B_s \to \mu^+\mu^-$, construct the relevant observables, and provide SM predictions. We focus in particular on uncertainties coming from the Cabibbo--Kobayashi--Maskawa (CKM) matrix elements. In Sec.~\ref{ch:extracting_C10}, we demonstrate that semileptonic $B_{(s)}$ decays can be used to determine the axial-vector NP contribution, with negligible impact of (pseudo)-scalar contributions. In Sec.~\ref{ch:section4}, we obtain our target regions for the $\adg$ and $\smumu$ observables. We conclude and give an outlook in Sec.~\ref{ch:conclusions}.

\section{Theoretical Framework for the \boldmath{$\bar B^0_s \to\mu^+\mu^-$} Decay}
\label{ch:theoretical_framework}

\subsection{Effective Hamiltonian}
In order to describe the $\bar B^0_s \to\mu^+\mu^-$ decay, we apply effective quantum field theory, where heavy SM and NP degrees of freedom are ``integrated out'', leading to a low-energy effective Hamiltonian of the following form \cite{Bobeth:2013uxa,DeBruyn:2012wk,Altmannshofer:2017wqy,Fleischer:2017yox}:
\begin{equation}\label{eq:hamil}
    \mathcal{H}_{\rm eff} = - \frac{G_{\rm F}}{\sqrt{2} \pi} V_{ts}^* V_{tb} \alpha_{\rm em} \left[ C_{10} O_{10} + C_{S} O_{S} + C_{P} O_{P} + C_{10}^\prime O_{10}^\prime + C_{S}^\prime O_{S}^\prime + C_{P}^\prime O_{P}^\prime \right] \ .
\end{equation}
Here, $G_{\rm F}$ is the Fermi constant, $V_{tb}$ and $V_{ts}^*$ are elements of the CKM matrix, and $\alpha_{\rm em}$ is the QED fine structure constant. The four-fermion operators\footnote{We note that our convention for $O_S^{(\prime)}$ and $O_S^{(\prime)}$ differs by a factor $m_b$ from \cite{Descotes-Genon:2015hea} but agrees with \cite{DeBruyn:2012wk,Altmannshofer:2017wqy,Fleischer:2017yox}.} are given as follows:
\begin{equation}
\begin{aligned}
    O_{10} &= (\bar s \gamma_\mu P_L b) (\bar \mu \gamma^\mu \gamma_5 \mu) \ , & O_{10}^\prime &= (\bar s \gamma_\mu P_R b) (\bar \mu \gamma^\mu \gamma_5 \mu) \ ,\\
    O_S &= m_b (\bar s P_R b) (\bar \mu \mu) \ , &
    O_S^\prime &= m_b (\bar s P_L b) (\bar \mu \mu) \ ,\\
    O_P &= m_b (\bar s P_R b) (\bar \mu \gamma_5 \mu) \ , &
    O_P^\prime &= m_b (\bar s P_L b) (\bar \mu \gamma_5 \mu) \ ,
\end{aligned}
\end{equation}
where $m_b$ is the bottom-quark mass, and $P_{R(L)} = \frac{1}{2}(1\pm \gamma_5)$. 
In general, the Wilson coefficients depend on the flavour of the final-state leptons \cite{Fleischer:2017ltw}. In this paper, we consider only decays into two muons and the corresponding coefficients. In the effective Hamiltonian describing the CP-conjugate processes, all CP-violating phases associated with the Wilson coefficients change their signs.

For convenience, we define the combinations 
\begin{align}\label{eq:pmdef}
    \mathcal{C}_i^\pm &\equiv C_i \pm C_i^\prime \ , \quad i \in \{10,S,P \} \ ,
\end{align}
for the axial-vector, scalar and pseudo-scalar operators. In the SM, only the axial-vector operator $O_{10}$ arises, and we introduce
    \begin{equation}
        C_{10} \equiv C_{10}^{\rm SM} + C_{10}^{\rm NP}\ .
    \end{equation} 

\subsection{Observables and Inputs}\label{sec:inputs}
Let us discuss the observables of $\bar B^0_s \to \mu^+ \mu^-$ in this subsection. We  use \eqref{eq:hamil} to calculate the decay amplitude through evaluating the hadronic matrix element between an initial $B^0_s$ meson and the final $\mu^+ \mu^-$ state. The decay constant $f_{B_s}$  encodes the hadronic physics and is known with impressive precision from Lattice QCD (LQCD) computations\footnote{Further hadronic parameters enter when consider also power-enhanced leading-logarithmic QED corrections \cite{Beneke:2019slt}, which play a minor role.}. The HFLAG average \cite{FlavourLatticeAveragingGroupFLAG:2021npn} utilising the results of \cite{Dowdall:2013tga,ETM:2016nbo,Hughes:2017spc,Bazavov:2017lyh} yields
\begin{equation}
    f_{B_s} = (230.3 \pm 1.3) \;\si{\mega \eV} \ .
\end{equation}
The branching ratio of this decay channel in the SM can be written as follows  \cite{Buras:2013uqa}:
\begin{equation}
\begin{aligned}
    {\mathcal{ B}}( B_s \to \mu^+ \mu^-) &= \frac{G_F^2 \abs{V_{ts} V_{tb}}^2 \alpha_{\rm em}^2 m_{B_s} f_{B_s}^2 \tau_{B_s}}{64 \pi^3} \sqrt{1 - \frac{4 m_\mu^2}{m_{B_s}^2}} 4 m_\mu^2 |C_{10}^{\rm SM}|^2 \bigg[
    \abs{P}^2
    + \abs{S}^2
    \bigg] \ ,
\end{aligned}
\label{eq:BR_Bsellell}
\end{equation}
which is a CP-averaged quantity. Alternatively, we may replace $\alpha_{\rm em}$ by 
\begin{equation}
    \alpha_{\rm em} = \frac{2 G_{\rm F}^2 m_W^4 \sin^4\theta_{\rm W}}{\pi^2}\ ,
\end{equation}
where $\theta_{\rm W}$ is the Weinberg angle. Numerically, we use $\alpha_{\rm em} = 1/132$, and for the Wilson coefficient we use
\begin{equation}
    C_{10}^{\rm SM}  
      = -4.194 \ ,
\end{equation} 
obtained in flavio \cite{Straub:2018kue} at the scale $\mu = m_b$.

The expression in \eqref{eq:BR_Bsellell}, which holds in the presence of interactions from beyond the SM, introduces the quantities \cite{DeBruyn:2012wk}
\begin{equation}
    P \equiv \frac{C_{10} - C_{10}^\prime}{C_{10}^{\rm SM}} + \frac{M_{B_s}^2}{2 m_\mu} \left( \frac{m_b}{m_b + m_s} \right) \left( \frac{C_P - C_P^\prime}{C_{10}^{\rm SM}} \right) \equiv \abs{P} e^{i \varphi_P} \ ,
    \label{eq:P}
\end{equation}
which depends on the combination $\cminus_{10}$ and 
\begin{equation}
    S \equiv \sqrt{1 - \frac{4 m_\mu^2}{M_{B_s}^2}} \frac{M_{B_s}^2}{2 m_\mu} \left( \frac{m_b}{m_b + m_s} \right) \left( \frac{C_S - C_S^\prime}{C_{10}^{\rm SM}} \right) \equiv \abs{S} e^{i \varphi_S} \ ,
    \label{eq:S}
\end{equation}
where $\varphi_P$ and $\varphi_S$ are CP-violating phases and we use $m_s= 0.093$ MeV for the strange quark mass in our numerical analysis. In the SM, we have by definition the following values:
\begin{equation}
    P_{\rm SM} =1 \ , \quad\quad S_{\rm SM}=0 \ .
\end{equation} 

The branching ratio  ${\mathcal{ B}}( B_s \to \mu^+ \mu^-)$ in (\ref{eq:BR_Bsellell}) exhibits helicity suppression, which is reflected by the $m_\mu^2$ factor. Looking at \eqref{eq:P} and \eqref{eq:S}, we observe that the helicity suppression can be lifted by the $\mathcal{C}_P^{(')}$ and $\mathcal{C}_S^{(')}$coefficients, entering with a factor $1/m_\mu$.

The expression of \eqref{eq:BR_Bsellell} defines the ``theoretical'' branching ratio at $t = 0$, where mixing effects are switched off. However, taking the $B^0_s$--$\bar{B}^0_s$ mixing effects properly into account, we measure the time-integrated branching ratio $\overline{\mathcal{B}} (B_s \to \mu^+ \mu^-)$, which is related to the theoretical one through \cite{DeBruyn:2012wk}
\begin{equation}
    \mathcal{B} (B_s \to \mu^+ \mu^-)_{\rm theo} = \left[\frac{1 - y_s^2}{1 + \mathcal{A}_{\Delta\Gamma_s} y_s}\right] \overline{\mathcal{B}} (B_s \to \mu^+ \mu^-) \ ,
\end{equation}
where
\begin{equation}
    \mathcal{B}(B_s \to \mu^+ \mu^-)_{\rm theo} = \frac{1}{2} \bigg[ \mathcal{B}(\bar B^0_s \to \mu^+ \mu^-) + \mathcal{B}(B^0_s \to \mu^+ \mu^-) \bigg] 
\end{equation} 
is the CP-averaged (``untagged'') theoretical branching ratio.  
The parameter  \cite{HFLAV} 
\begin{equation}
    y_s \equiv \frac{\Gamma_{\rm L}^{(s)}-\Gamma_{\rm H}^{(s)}}{\Gamma_{\rm L}^{(s)}+\Gamma_{\rm H}^{(s)}} = \frac{\Delta \Gamma_s}{ 2 \Gamma_s} = 0.062 \pm 0.004
\end{equation}
depends on the decay width difference $\Delta \Gamma_s$ between the $B_s$ mass eigenstates and on the $B_s$ lifetime ${\tau_{B_s}}=1/\Gamma_s $. Due to the sizeable $\Delta \Gamma_s$, we gain access to
\begin{equation}
    \mathcal{A}_{\Delta\Gamma_s} = \frac{\abs{P}^2 \cos (2 \varphi_{P} - \phi_s^{\rm NP}) - \abs{S}^2 \cos (2 \varphi_{S} - \phi_s^{\rm NP})}{\abs{P}^2  + \abs{S}^2} \ .
    \label{eq:ADG}
\end{equation}
This observable is given by $+1$ in the SM but can generally take any value between $-1$ and $+1$. The phase $\phi_s^{\rm NP}$ is related to CP-violating contributions from beyond the SM to $B^0_s$--$\bar B^0_s$ mixing:
\begin{equation}
\label{eq:phisNP_def}
 \phi_s^{\rm NP} =   \phi_s - \phi_s^{\rm SM} \ .
\end{equation}
Following \cite{Barel:2020jvf,DeBruyn:2022zhw}, we combine the value of $\phi_s$ with the SM prediction $\phi_s^{\rm SM} $.  The $\phi_s$ comes from decays of the kind $B^0_s \to J/\psi \phi$, where contributions from penguin topologies have been taken into account \cite{Barel:2022wfr}:
\begin{equation}
    \phi_s =   (-4.2 \pm 1.4)^\circ \ .
\end{equation}
The SM value arises from an analysis of the Unitarity Triangle (UT) utilising only information from the angle $\gamma$ and the side $R_b$ \cite{Barel:2022wfr}:
\begin{equation}
    \phi_s^{\rm SM}  =   (-2.01 \pm 0.12)^\circ \ .
\end{equation}
Finally, using \eqref{eq:phisNP_def}, the NP phase is determined as \cite{Barel:2022wfr}
\begin{equation}\label{eq:phisNP}
\phi_s^{\rm NP}=(2.2 \pm 1.4)^\circ \ ,
\end{equation}
which is the value we use in order to include NP contributions to $B^0_s$--$\bar B^0_s$ mixing.

We note that $\adg$ can be obtained from measurements of the $B_s \to \mu^+\mu^-$ effective lifetime, defined as follows (see also \cite{DeBruyn:2012wk,Fleischer:2017yox}):
\begin{equation}
     \tau_{\mu\mu}^s \equiv \frac{\int_{0}^{\infty} t \langle \Gamma (B_{s}(t) \to \mu^+ \mu^- )\rangle dt}{\int \int_{0}^{\infty}  \langle \Gamma (B_{s}(t) \to \mu^+ \mu^- )\rangle dt}  \ .
\end{equation}
Measuring this quantity requires time information for untagged data samples. ATLAS \cite{ATLAS:2023trk}, CMS  \cite{CMS:2022mgd} and the LHCb Collaboration \cite{LHCb:2021awg} have all recently reported measurements of the effective lifetime $\tau_{\mu\mu}^s$:
\begin{equation}
\begin{aligned}
    \tau_{\mu\mu}^s|_{\rm LHCb \; 2021} &= (2.07 \pm 0.29) \; \si{\pico s} \ ,\\
    \tau_{\mu\mu}^s|_{\rm CMS \; 2022} &= (1.83\pm 0.23) \;  \si{\pico s} \ ,\\
   \tau_{\mu\mu}^s|_{\rm ATLAS \; 2023} &= (0.99 \pm 0.45) \; \si{\pico s} \ .\\
\end{aligned}
\end{equation}
Using the following relation \cite{DeBruyn:2012wk}:
\begin{equation}\label{eq:adg}
 \adg = \frac{1}{ y_s} \left[ \frac{(1-y_s^2)\tau_{\mu\mu}^s-(1+
 y_s^2)\tau_{B_s}}{2\tau_{B_s}-(1-y_s^2)\tau_{\mu\mu}^s} \right] \ ,
\end{equation} 
the above results can be translated into bounds on $\adg$, 
yielding:
\begin{equation}
\begin{aligned}
    \adg|_{\rm LHCb \; 2021} &= 8.44 \pm 7.11 \ , \\
    \adg|_{\rm CMS \; 2022} &= 3.75 \pm 3.68 \ , \\
    \adg|_{\rm ATLAS \; 2023} &= -4.08 \pm 2.52 \ . \\
\end{aligned}
\label{eq:ADG_exp}
\end{equation}
Unfortunately, the uncertainties are still too large to constrain the observable within its model-independent range of $-1 \leq \adg \leq 1$. 

Finally, we can also consider the full time-dependent rate of the $B_s$ mesons decaying into two muons with helicity $\lambda = L,R$. Even though the expression in \eqref{eq:BR_Bsellell} corresponds to a summation over helicities, it is now useful to work on the case for specific helicities. A flavour-tagged analysis would then give access to the CP-violating decay rate asymmetry \cite{DeBruyn:2012wk,Buras:2013uqa}:
\begin{equation}
    \frac{\Gamma(B^0_s(t) \to \mu_\lambda^+ \mu_\lambda^-) - \Gamma(\bar B^0_s(t) \to \mu_\lambda^+ \mu_\lambda^-)}{\Gamma(B^0_s(t) \to \mu_\lambda^+ \mu_\lambda^-) + \Gamma(\bar B^0_s(t) \to \mu_\lambda^+ \mu_\lambda^-)} = \frac{\mathcal{C}_{\mu\mu}^\lambda \cos (\Delta M_s t) + \smumu \sin (\Delta M_s t)}{\cosh (y_s t/\tau_{B_s}) + \adg \sinh (y_s t/\tau_{B_s})} \ ,
    \label{eq:time_dep_helicities}
\end{equation}
with the CP asymmetries
\begin{equation}
    \mathcal{C}_{\mu\mu}^\lambda = -\eta_\lambda \left[ \frac{2 \abs{P S} \cos(\varphi_P - \varphi_S)}{\abs{P}^2 + \abs{S}^2} \right] \equiv -\eta_\lambda \cmumu \ ,
    \label{eq:Cmumu}
\end{equation}
\begin{equation}
    \smumu = \frac{\abs{P}^2 \sin(2 \varphi_P - \phi_s^{\rm NP}) - \abs{S}^2 \sin(2 \varphi_S - \phi_s^{\rm NP})}{\abs{P}^2 + \abs{S}^2} \ ,
    \label{eq:Smumu}
\end{equation}
both of which are zero in the SM. The observable $\mathcal{C}_{\mu\mu}^\lambda$ depends on the helicity of the lepton pair ($\eta_L = +1$ and $\eta_R = -1$), making it difficult to measure. However, we stress that even knowing its sign would already be useful to resolve discrete ambiguities that arise when determining (pseudo)-scalar Wilson coefficients \cite{Fleischer:2017yox}. We note that no data are yet available for $\smumu$ and $\cmumu$. 

Keeping the time dependence but summing over the muon helicities, \eqref{eq:time_dep_helicities} becomes
\begin{equation}\label{eq:timedep}
    \frac{\Gamma(B^0_s(t) \to \mu^+ \mu^-) - \Gamma(\bar B^0_s(t) \to \mu^+ \mu^-)}{\Gamma(B^0_s(t) \to \mu^+ \mu^-) + \Gamma(\bar B^0_s(t) \to \mu^+ \mu^-)} = \frac{\smumu \sin (\Delta M_s t)}{\cosh (y_s t/\tau_{B_s}) + \adg \sinh (y_s t/\tau_{B_s})} \ .
\end{equation}

The three observables $\adg$, $\smumu$, and $\cmumu$ are theoretically clean since the decay constant $f_{B_s}$ cancels in all of them. They satisfy the relation
\begin{equation}
    (\adg)^2 + (\smumu)^2 + (\cmumu)^2 = 1 \ ,
    \label{eq:sphere}
\end{equation}
and are therefore not independent of one another.

\subsection{Information from Branching Ratios}
The SM prediction for the branching ratio in \eqref{eq:BR_Bsellell} depends on the CKM matrix elements entering $|V_{ts} V_{tb}|$. Using the Wolfenstein parametrization, we obtain
\begin{align}
    |V_{ts}V_{tb}|
 = \phantom{\lambda}|V_{cb}|\left[1-\frac{\lambda^2}{2} \left(1-2 R_b \cos \gamma\right)\right] + \mathcal{O}\left(\lambda^6\right)\:,
 \label{eq:vtsvtb}
\end{align}
where $\lambda \equiv |V_{us}| = 0.22$ \cite{ParticleDataGroup:2022pth}, the parameter
\begin{equation}
  R_b \equiv \left| \frac{V_{ud}V_{ub}^*}{V_{cd}V_{cb}^*} \right| = \left(1-\frac{\lambda^2}{2}\right)\frac{1}{\lambda}\left|\frac{V_{ub}}{V_{cb}}\right| \ ,
\end{equation}
is the usual side of the UT from the origin to the apex,
and $\gamma$ is the corresponding angle. 
Expression~\eqref{eq:vtsvtb} is governed by $|V_{cb}|$ up to $\mathcal{O}({\lambda^2})$ corrections, which involve $R_b$ and $\gamma$.

To determine $|V_{ts}V_{tb}|$, we thus require values for $|V_{ub}|$ and $|V_{cb}|$. This poses a problem, as these CKM factors carry a substantial, hidden theoretical uncertainty. There are two methods to determine $|V_{ub}|$ and $|V_{cb}|$. One uses exclusive decay processes and the other inclusive measurements. Ideally, both determinations should agree with each other, but in practice, tensions arise. These long-standing tensions constitute the key uncertainty for the branching ratio, and special care is needed. Therefore, we give separately the inclusive/hybrid and exclusive cases studied in \cite{DeBruyn:2022zhw}. 
The hybrid approach combines the exclusive $|V_{ub}|$ with the inclusive $|V_{cb}|$ values. From \eqref{eq:vtsvtb} we see that $|V_{ub}|$ only enters via higher order corrections. Therefore, the inclusive and hybrid scenarios for $|V_{ts}V_{tb}|$ coincide within the precision considered here. Using input from inclusive $b \to c \ell \nu_\ell$ decays \cite{Bordone:2021oof}, we obtain:
\begin{equation}\label{eq:hybrid}
|V_{ts}V_{tb}|_{\rm incl/hybrid} = (41.4 \pm 0.5)\times 10^{-3} \ .
\end{equation}
Taking the exclusive $|V_{cb}|$ from the HFLAV collaboration \cite{HFLAV:2019otj}, we find
\begin{equation}
|V_{ts}V_{tb}|_{\rm excl} = (38.4 \pm 0.5)\times 10^{-3},
\end{equation}
which differs from \eqref{eq:hybrid} at the $4\sigma$ level due to the different $|V_{cb}|$ values.  
Finally, we find the following predictions for the SM values of the branching ratios:
\begin{align}
   \overline{\mathcal{B}}(B_s \to \mu^+\mu^-)_{\rm SM}^{\rm incl/hybrid} &= (3.71 \pm 0.04|_{f_{B_s}} \pm 0.09|_{\rm CKM} \pm 0.02|_{\rm source}) \cdot 10^{-9} \nonumber\\ 
   & =(3.71 \pm 0.10) \cdot 10^{-9} 
    \ ,
    \label{eq:RSM1}
\end{align}
\begin{align}
 \overline{\mathcal{B}}(B_s \to \mu^+\mu^-)_{\rm SM}^{\rm excl } &= (3.19 \pm 0.04|_{f_{B_s}} \pm 0.08|_{\rm CKM} \pm 0.02|_{\rm source}) \cdot10^{-9} \nonumber \\
 & = (3.19 \pm 0.09) \cdot 10^{-9} 
    \ .
\label{eq:barRSM1}
\end{align}
It becomes clear that there is a significant variation depending on the CKM parametrization. Thus, special attention is needed for the CKM uncertainties. The results in \eqref{eq:RSM1} and \eqref{eq:barRSM1} are in the ballpark of other theoretical predictions \cite{Beneke:2019slt}, and we note that the spread between inclusive and exclusive results is much wider than the quoted uncertainties.

The current experimental world average is given as follows \cite{HFLAV:2022pwe}:
\begin{equation}
    \overline{\mathcal{B}}(B_s \to \mu^+\mu^-)=(3.45 \pm 0.29)\times 10^{-9} \ ,
    \label{eq:Bsmumu_BR_HFLAV}
\end{equation}
relying mainly on recent measurements by LHCb \cite{LHCb:2021awg}, CMS \cite{CMS:2022mgd}, and ATLAS \cite{ATLAS:2018cur}.  
This experimental average is smaller than the inclusive SM prediction in \eqref{eq:RSM1} but larger than the exclusive SM prediction in \eqref{eq:barRSM1}.

Moving now towards studies of NP, we introduce the ratio between the experimental and SM branching ratios:
\begin{equation}
    \overline R = \frac{\overline{\mathcal{B}}(B_s \to \mu^+\mu^-)}{\overline{\mathcal{B}}(B_s \to \mu^+\mu^-)_{\rm SM}}
    \ .
    \label{eq:barRexp}
\end{equation}
In the SM, by definition $\overline{R} = 1$. Combining the current experimental results in \eqref{eq:Bsmumu_BR_HFLAV} with the  inclusive/hybrid and exclusive values for the SM predictions in \eqref{eq:RSM1} and \eqref{eq:barRSM1}, we obtain
\begin{align}
    \overline R_{\rm incl/hybrid} &=0.93 \pm 0.08 \ ,  \\
    \overline R_{\rm excl} &= 1.08 \pm 0.10 \ ,
\end{align}
where we see again that the experimental result can be either larger or smaller than the SM prediction, depending on the choice of the CKM parameterization.

\begin{figure}
   \centering
    \includegraphics[width = 0.7\textwidth]{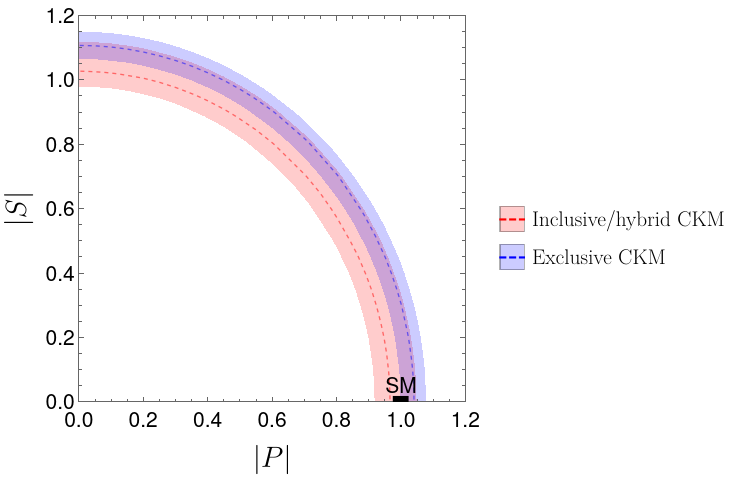}
    \caption{Constraints in the $|P|$--$|S|$ plane from the current world average experimental measurement of $\overline{\mathcal{B}}(B_s \to \mu^+\mu^-)$ for the inclusive/hybrid and exclusive CKM parameterizations.}
    \label{fig:P-S_plots}
\end{figure}

Taking into account the difference between the time-integrated and theoretical branching ratios, $\overline{R}$ can be written as follows \cite{Buras:2013uqa}:
\begin{align}
    \overline R
   & = \bigg[ \frac{1 + \adg y_s}{1 + y_s}\bigg] (|P|^2 + |S|^2)
   \nonumber  \\
    &=\left[\frac{1+y_s \cos(2\varphi_P-\phi_s^{\rm NP})}{1+y_s}\right] |P|^2 + \left[\frac{1-y_s \cos(2\varphi_S - \phi_s^{\rm NP})}{1+y_s}\right] |S|^2 \ ,
    \label{eq:barR} 
\end{align}
where $S$ and $P$ are given in \eqref{eq:S} and \eqref{eq:P}, respectively. Using the value of $\phi_s^{\rm NP}$ in \eqref{eq:phisNP} and assuming real (pseudo)-scalar couplings, i.e.~$\varphi_P = \varphi_S=0$, this ratio constrains a circle in the $|P|$--$|S|$ plane. In Fig.~\ref{fig:P-S_plots}, we show the circles which result from using the inclusive/hybrid and exclusive values of $\overline{R}$. We note that within their respective uncertainties these circles overlap. The SM point, with $P_{\rm SM}=1$ and $S_{\rm SM}=0$, is also indicated and lies between both predictions. Allowing for physics beyond the SM, entering through $\cminus_{10}$ and/or through new (pseudo)-scalar couplings in $\cminus_{S,P}$, the branching ratio thus constrains a combination of these possible new couplings. If present, these interactions have to be disentangled from $\cminus_{10}$. In the following, we discuss how to do so. 
For our analysis, we will use the inclusive/hybrid value for the CKM elements.

\section{\boldmath Input from Semileptonic Rare $B_{(s)}$ Decays}
\label{ch:section3}
\label{ch:extracting_C10}
The leptonic $B_s \to \mu^+\mu^-$ decay is governed by combinations of the axial-vector, scalar and pseudo-scalar Wilson coefficients. The current experimental value of the $B_s\to \mu^+\mu^-$ branching ratio, which is in agreement with the SM prediction within the uncertainties, seems to indicate marginal space for NP in $\cminus_{10}$ (see e.g.~\cite{CMS:2022dbz,Alguero:2023jeh}). However, this only holds if (pseudo)-scalar new interactions are neglected. We will demonstrate that the constraints on such effects are rather weak, indicating that possible new (pseudo)-scalar interactions actually cannot be excluded. As seen in \eqref{eq:P} and \eqref{eq:S}, $\cminus_P$ and $\cminus_S$ enter in $P$ and $S$, respectively. In addition, $P$ contains the coefficient $\cminus_{10}$. The quantities $P$ and $S$ can be determined through observables of the $B_s \to \mu^+\mu^-$ channel, as discussed in \cite{Fleischer:2017ltw,Fleischer:2017yox}. However, in order to constrain the corresponding Wilson coefficients, we have to disentangle the $\cminus_{10}$ and $\cminus_P$ contributions which both enter $P$. Since our goal is to probe possible NP effects entering through (pseudo)-scalar contributions to $B_s\to \mu^+\mu^-$, we thus require external input for $\cminus_{10}$. In this section, we will show that (pseudo)-scalar couplings have a negligible effect on the semileptonic rare decays. Therefore, $\cminus_{10}$ can be extracted independently of the (pseudo)-scalar coefficients from semileptonic decays of the kind $B\to K^{(*)} \mu^+\mu^-$ and $B_s\to \phi\mu^+\mu^-$.

\subsection{\boldmath Theoretical Framework}
\begin{table}
  \centering
  \begin{tabular}{ccc}
    \toprule
    Decay & Wilson Coefficients \\
       \midrule
  $B_s\to \phi \mu^+\mu^-$  & $\cplus_9, \cminus_9, \cplus_{10}, \cminus_{10}, (\cminus_P, \cminus_S)$ \\
   $B\to K^* \mu^+\mu^-$ &  $\cplus_9, \cminus_9, \cplus_{10}, \cminus_{10}, (\cminus_P, \cminus_S)$ \\
  $B\to K \mu^+\mu^-$  & $\cplus_9, \cplus_{10}, (\cplus_P, \cplus_S)$ \\
  $B_s\to  \mu^+\mu^-$ &  $\cminus_{10}, \cminus_P, \cminus_S$ \\
    \bottomrule
  \end{tabular}
  \caption{Rare $B$ decays and their dependence on the Wilson coefficients. Parentheses indicate negligible dependencies.}
  \label{tab:wilson}
\end{table}
The semileptonic $B_{(s)} \to K^{\ast}(\phi) \mu^+\mu^-$ and $B \to K\mu^+\mu^-$ modes are described by the low-energy effective Hamiltonian
\begin{equation}\label{eq:ham}
    \mathcal{H}_{\rm eff} = - \frac{G_F}{\sqrt{2}\pi} V_{ts}^*V_{tb} \alpha_{\rm em} \sum\limits_{i \in I} C_i \mathcal{O}_i  \ ,
\end{equation}
where we have neglected doubly Cabibbo-suppressed terms. The sum runs over the index $I = \{1c, 2c, 3, 4, 5, 6, 8, 7^{(\prime)}, 9^{(\prime)}\mu, 10^{(\prime)}\mu, S^{(\prime)}\mu, P^{(\prime)}\mu, T^{(\prime)}\mu\}$ and includes all operators that can contribute to the modes. In particular, in contrast with $B_s \to \mu^+\mu^-$, now also the following operators enter: 
\begin{equation}
 \mathcal{O}_{9^{(\prime)}\mu} = [\bar s \gamma^\mu P_{L(R)} b] (\bar \mu \gamma_\mu \mu) \ ,  \quad        \mathcal{O}_{7^{(\prime)}} = \frac{1}{e} m_b [\bar s \sigma^{\mu\nu} P_{R(L)} b] F_{\mu\nu} \ ,
\end{equation}
with $\sigma_{\mu\nu} = \frac{i}{2}[\gamma_\mu, \gamma_\nu]$. In the SM, we have \cite{Straub:2018kue}
\begin{equation}
    C_9^{\rm SM} = 4.114 \ , \quad\quad C_7^{\rm SM} = -0.296 \ .
\end{equation}
Expressions for the differential rate of the $B_{(s)} \to K^*(\phi) \mu^+\mu^-$ modes are given in e.g.~\cite{Descotes-Genon:2015hea}, while those for $B\to K\mu^+\mu^-$ can be found in e.g.~\cite{Fleischer:2022klb}. In the following, we neglect NP contributions to $C_7$ and the tensor coefficient $C_T$. The branching ratios then depend on specific combinations of $C_{9,10,S,P}$, given in Table \ref{tab:wilson}, where we have defined
\begin{equation}\label{eq:C9def}
    \mathcal{C}_9^\pm \equiv C_9 \pm C_9^\prime
\end{equation}
in analogy with \eqref{eq:pmdef}. 

In the $B_{(s)}\to K^*(\phi)$ modes, there are several additional observables besides the branching ratio (see, e.g., \cite{Descotes-Genon:2015hea} for an overview). It is important to note that, unlike the $B_s \to \mu^+\mu^-$ channel, the semileptonic modes are not helicity suppressed. As a result, they are much less sensitive than the $B_s \to \mu^+\mu^-$ decay to $C_P$ and $C_S$.

Recent predictions for the branching ratio of $B\to K\mu^+\mu^-$ can be found in \cite{Fleischer:2022klb}. For the $B_s\to \phi$ and $B\to K^*$ modes, we present the SM predictions using the inputs given above, the LQCD and Light-cone sum rule (LCSR) form factors from \cite{Bharucha:2015bzk} and perturbative long-distance effects from \cite{Descotes-Genon:2015hea}. We find the following SM predictions:
\begin{align}
    \overline{\mathcal{B}}(B_s \to \phi \mu^+ \mu^- )|_{\rm incl/hybrid} {}&= (5.63\pm 0.58) \cdot 10^{-8}\label{eq:smpred_bsphimumu} \ , \\
       \overline{\mathcal{B}}(B_s \to \phi \mu^+ \mu^- )|_{\rm excl} {}&= (4.85\pm 0.49) \cdot 10^{-8} \ ,
\end{align}
where we have integrated the differential rate over the $1.1<~q^2<~6.0 \; \si{\giga \eV^2}$ range, with $q^2$ denoting the four-momentum transfer to the $\mu^+\mu^-$ pair. Analogous to the discussion in Sec.~\ref{sec:inputs}, we converted the theoretical rate to the experimental branching ratio corresponding to time-integrated untagged data samples following \cite{Descotes-Genon:2015hea}. 
Again we note the significant difference in the SM predictions depending on the CKM matrix element determinations, which was also noted, e.g., in \cite{Fleischer:2022klb}. For the $K^*$ modes, we find
\begin{align}
     \mathcal{B}(B\to K^{*\pm} \mu^+ \mu^- )|_{\rm incl/hybrid} {}&= (5.35\pm 0.72) \cdot 10^{-8} \nonumber  \ ,\\
      \mathcal{B}(B\to  K^{*\pm}  \mu^+ \mu^- )|_{\rm excl} {}&= (4.61\pm 0.61) \cdot 10^{-8} \ ,
\end{align}
and
\begin{align}
    \overline{\mathcal{B}}(B\to  K^{*0}  \mu^+ \mu^- )|_{\rm incl/hybrid} {}&= (5.00\pm 0.64) \cdot 10^{-8} \nonumber \ ,\\
       \overline{\mathcal{B}}(B\to K^{*0} \mu^+ \mu^- )|_{\rm excl} {}&= (4.31\pm 0.55) \cdot 10^{-8} \ .
\end{align}
These results agree within uncertainties with \cite{Gubernari:2022hxn,Bharucha:2015bzk,Straub:2018kue}. The experimental values measured by the LHCb Collaboration \cite{LHCb:2021zwz,LHCb:2014cxe} are given as follows:
\begin{equation}
    \overline{\mathcal{B}}(B_s\to \phi \mu^+ \mu^- )|_{\rm exp} = (2.88 \pm 0.21) \cdot 10^{-8}
 \ ,\end{equation}
and 
\begin{equation}
    \overline{\mathcal{B}}(B^\pm \to K^{*\pm}\mu^+\mu^-)|_{\rm exp} = (3.66 \pm 0.87) \cdot 10^{-8} \ .
\end{equation}
We find that the experimental branching ratios are smaller than the SM predictions, a trend also reflected in the semileptonic $B \to K\mu^+\mu^-$ channel (see, e.g.,~\cite{Fleischer:2022klb}). These anomalies are well known and could be due to NP in (axial)-vector operators (see, e.g., \cite{Bobeth:2017vxj,Alok:2017sui, Datta:2019zca,Altmannshofer:2021qrr,Alguero:2022wkd,Gubernari:2022hxn,Carvunis:2021jga,Mahmoudi:2022hzx,SinghChundawat:2022zdf,Alguero:2023jeh}).

\subsection{On Extracting \boldmath{$\cminus_{10}$}}
The $\cminus_{10}$ coefficient enters the rare semileptonic modes $B_{(s)} \to K^*(\phi) \mu^+\mu^-$, but the (pseudo)-scalar coefficients $\cminus_P$ and $\cminus_S$ also enter these modes. We will show that the impact of $\cminus_{P,S}$ on the semileptonic decays is actually negligible.

\begin{figure}
    \centering
      \subfloat[]{\includegraphics[height=0.35\textwidth]{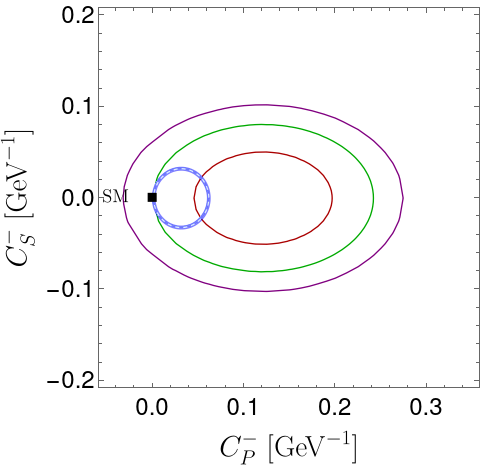}}
      \hfill
    \subfloat[Zoomed in 5 times]{\includegraphics[height=0.35\textwidth]{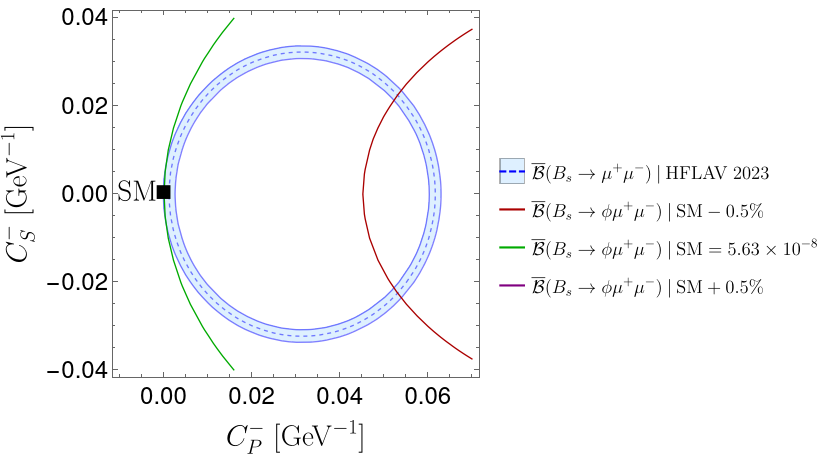}}
      \caption{Correlation between $\cminus_P$ and $\cminus_S$ for $\mathcal{C}_{10}^{- \rm NP} = 0$. The blue circular band is the $1\sigma$ constraint from $\overline{\mathcal{B}}(B_s \to \mu^+\mu^-$) in \eqref{eq:Bsmumu_BR_HFLAV}, and the ellipses indicate values of $\overline{\mathcal{B}}(B_s \to \phi\mu^+\mu^-)$ in the $q^2 \in [1.1,6.0] \; \si{\giga\eV^2}$ bin.}
    \label{fig:CPCSplot_C10_0}
\end{figure}

We illustrate in Fig.~\ref{fig:CPCSplot_C10_0} the current constraint from the branching ratio of $B_s \to \mu^+\mu^-$ in the $\cminus_P$--$\cminus_S$ plane, assuming the coefficients to be real. In addition, we show the branching ratio of $B_s \to \phi\mu^+\mu^-$ as a function of $\cminus_P$ and $\cminus_S$. The three ellipses represent the SM value in \eqref{eq:smpred_bsphimumu} and the SM value $\pm$ 0.5 \%, specifically $[5.60,5.66] \times 10^{-8}$. For Fig.~\ref{fig:CPCSplot_C10_0}, we assume $\mathcal{C}_{10}^{- \rm NP} = 0$, but we find similar results when $\mathcal{C}_{10}^{- \rm NP} \neq 0$. We observe that the current $B_s \to \mu^+\mu^-$ measurement already constrains $\abs{\cminus_{P,S}} \leq 0.06 \; \si{\giga \eV^{-1}}$. On the other hand, we find that a NP contribution of that size affects the semileptonic branching ratio only at the percent level. This demonstrates that the leptonic decay is indeed significantly more sensitive than the semileptonic modes to (pseudo)-scalar NP and that $\cminus_{P,S}$ can only affect the branching ratio of $B_s \to \phi\mu^+\mu^-$ at the level of less than one percent. We obtain similar results for the $B^\pm \to K^{\pm \ast} \mu^+\mu^-$ and $B^\pm  \to K^\pm \mu^+\mu^-$ decays.

The small sensitivity of the semileptonic decays with respect to new (pseudo)-scalar contributions can also be seen by looking at the observables $R_S$ and $R_W$, defined in (66) and (68) of \cite{Descotes-Genon:2020tnz}. These are constructed to be sensitive to scalar NP contributions, and $R_W$ vanishes for $m_\ell = 0$ and $\cminus_S = \cminus_T = 0$. However, as was also noted in \cite{Descotes-Genon:2020tnz}, $R_S$ only changes minimally when allowing for new scalar contributions. In addition, $R_W$ does get affected by new scalar couplings, but its value remains of $\mathcal{O}(10^{-2})$. Therefore, only very precise experimental information on these observables, which are currently not measured, might help in constraining $\cminus_{P,S}$. 

We have demonstrated that semileptonic rare $B$ decays are much less sensitive to (pseudo)-scalar NP than the leptonic $B_s \to \mu^+\mu^-$ channel. For this reason, the semileptonic decays can be used to extract $\cminus_{9}$ and $\cminus_{10}$ with negligible impact from (pseudo)-scalar NP. This requires considering all the available data in the semileptonic rare decays, including angular coefficients and possible future CP asymmetry measurements. This would then allow for the determination of the complex $C_{10}^{(')}$ and $C_{9}^{(')}$ coefficients along the lines of \cite{Fleischer:2022klb}. The goal of our current paper is not to perform such an analysis. Therefore, in the following we consider benchmark scenarios for $C_{10}^{(')}$. Most importantly, we stress that by the time CP asymmetries in $B_s\to \mu^+\mu^-$ are measured, the complex (axial)-vector current Wilson coefficient can be extracted from the semileptonic rare decays.

\section{\boldmath{Target Regions for New CP Violation in the (Pseudo)-Scalar Sector 
}} 
\label{ch:section4}
\subsection{Utilizing the Branching Ratio and $\adg$}

\begin{figure}
    \centering
    \subfloat[$\mathcal{C}_{10}^{- \rm NP}= 0.30\abs{C_{10}^{\rm SM}}$]{\includegraphics[height=0.38\textwidth]{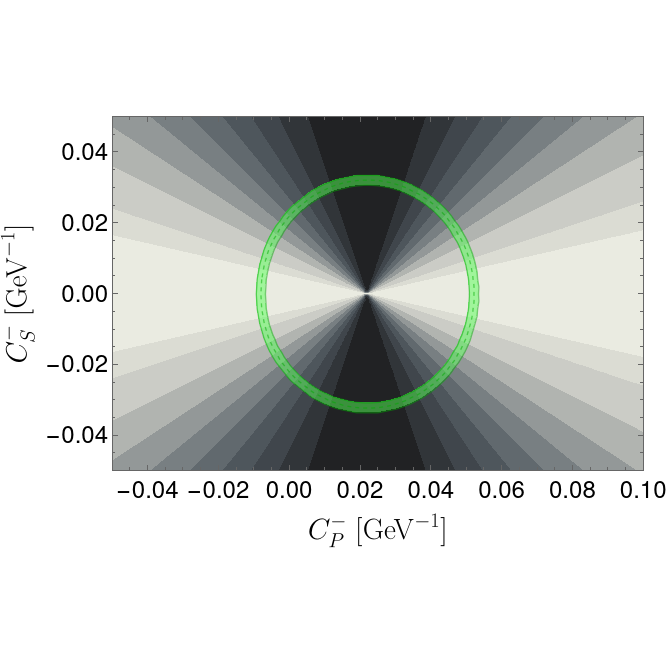} \label{fig:adg_plot_realandcomplex_c10_a}}
    \hfill
    \subfloat[$\mathcal{C}_{10}^{- \rm NP}= 0.30\abs{C_{10}^{\rm SM}} e^{i \, 60^\circ }$]{\includegraphics[height=0.38\textwidth]{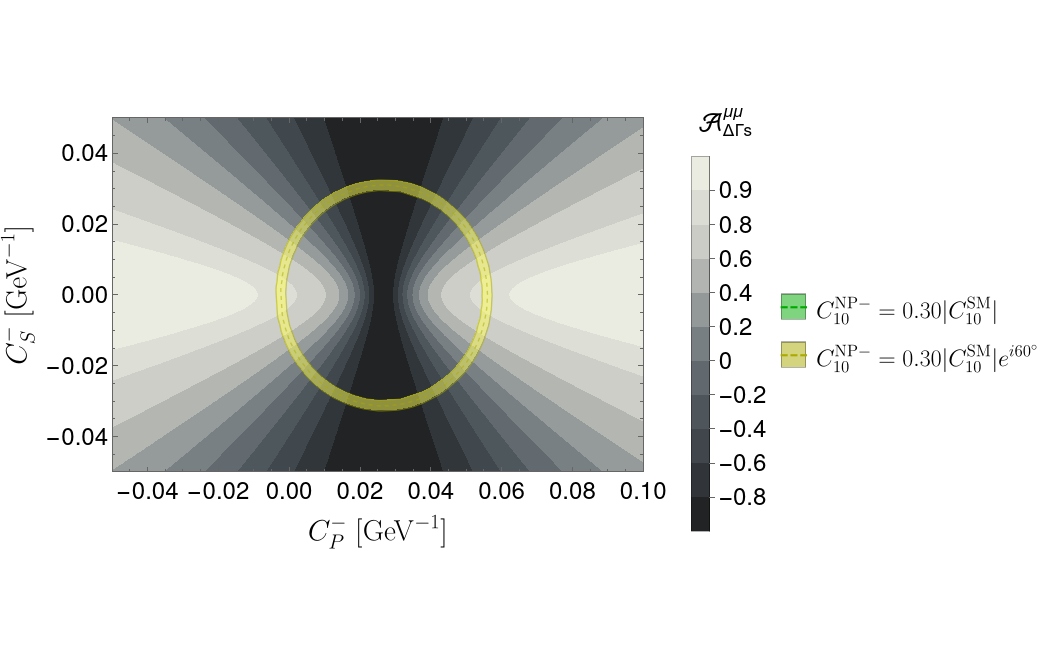} \label{fig:adg_plot_realandcomplex_c10_b}}
    \caption{Constraints from $\adg$ in the $\cminus_P$--$\cminus_S$ plane for a real (a) and complex (b) $\cminus_{10}$. The circular bands show the $1\sigma$ regions consistent with the measurement of $\overline{\mathcal{B}}(B_s \to \mu^+\mu^-)$ using the inclusive/hybrid CKM values in \eqref{eq:hybrid}.}
    \label{fig:adg_plot_realandcomplex_c10}
\end{figure}

Having obtained $\cminus_{10}$ through semileptonic rare $B_{(s)}$ decays, we can use the branching ratio and $\adg$ of the $B_s~\to~\mu^+\mu^-$ mode to probe (pseudo)-scalar couplings, assuming both are real. 
As shown in Fig.~\ref{fig:P-S_plots}, the branching ratio fixes a circular relation between $S$ and $P$. 
Using $\cminus_{10}$ as input, we can convert this to constraints on $\cminus_P$ and $\cminus_S$. 
To illustrate this, we consider both a real and complex benchmark scenario as input. Clearly, the latter scenario presents an exciting situation, since a new source of CP violation would already have been established. In this case, it would be important to explore whether also new (pseudo)-scalar interactions are present. The benchmark scenarios we consider here are merely to illustrate our strategy and to show the effect of complex phases in $C_{10}^-$ on the observables in $B_s\to \mu^+\mu^-$. Because the $C_{9}^{(')}$ coefficient does not enter our analysis of $B_s\to \mu^+\mu^-$, there is a lot of freedom to accommodate the (future) rare semileptonic measurements. As such, we consider NP effects of $30\%$ of the Standard Model value
\begin{align}
\mathcal{C}_{10}^{- \rm NP} &= 0.30\abs{C_{10}^{\rm SM}}\label{eq:c10minus_benchmarkreal} \ , \\
\mathcal{C}_{10}^{-\rm NP} &= 0.30\abs{C_{10}^{\rm SM}} e^{i \: 60^\circ} \label{eq:c10minus_benchmark}\ .
\end{align}
In Fig.~\ref{fig:adg_plot_realandcomplex_c10}, we show the constraints in the $\cminus_P$--$\cminus_S$ plane for both benchmarks using the inclusive/hybrid scenario for the CKM factors in \eqref{eq:hybrid} and the input parameters in Sec.~\ref{sec:inputs}. The circular bands show the $1\sigma$ regions consistent with the current measurement of the branching ratio of $B_s \to \mu^+\mu^-$ in \eqref{eq:Bsmumu_BR_HFLAV}. 

The (pseudo)-scalar couplings can be further constrained by employing $\adg$. For the real benchmark in \eqref{eq:c10minus_benchmarkreal}, we show in Fig.~\ref{fig:adg_plot_realandcomplex_c10_a} the values of $\adg$ in the $\cminus_P$--$\cminus_S$ plane corresponding to straight lines (as already shown in \cite{Fleischer:2017ltw}). Taken together with the circular band from $\overline{\mathcal{B}}(B_s \to \mu^+\mu^-)$, a measurement of $\adg$ thus lets us 
determine $\cminus_P$ and $\cminus_S$ up to a fourfold ambiguity.

For the complex benchmark in \eqref{eq:c10minus_benchmark}, $\adg$ gives curved contours in the $\cminus_P$--$\cminus_S$ plane as shown in Fig.~\ref{fig:adg_plot_realandcomplex_c10_b}. In this case the point $\cminus_{S,P}=0$ is allowed because our benchmark point for $\cminus_{10}$ accommodates the current experimental branching ratio of $B_s\to \mu^+\mu^-$. Analogous to before, the intersection between the branching ratio and $\adg$ determines $\cminus_P$ and $\cminus_S$. However, in this case, an interesting new feature arises as certain values of $\adg$ are now excluded. For the current branching ratio measurement and our benchmark scenario on \eqref{eq:c10minus_benchmark}, specifically, we find $|\adg| < 0.87$. Turning things around, with only real (pseudo)-scalar couplings, we can thus obtain a bound on $\adg$ from $\overline{\mathcal{B}}(B_s \to \mu^+\mu^-)$ using \eqref{eq:ADG}. 

At the moment, we cannot constrain these contributions any further, although pioneering measurements of $\adg$ are already available through the $B_s \to \mu^+\mu^-$ effective lifetime \eqref{eq:ADG_exp}. It will be interesting to see how future measurements of this observable will develop.

\subsection{\boldmath Utilizing Mixing-Induced CP Violation}
Additional information on possible new sources of CP violation comes from the mixing-induced CP asymmetry $\smumu$ in \eqref{eq:Smumu}. No measurements of $\smumu$ are currently available; extracting it would require a time-dependent, flavour-tagged analysis as can be seen in \eqref{eq:timedep}. 

We first consider real $\cminus_{10}, \;\cminus_{P}$ and $\cminus_S$. In that case, we can only have a non-zero $\smumu$ due to $\phi_s^{\rm NP}$ entering in \eqref{eq:S}. We then have
\begin{equation}
    \smumu = - \sin\phi_s^{\rm NP} \left(\frac{\abs{P}^2 - \abs{S}^2}{\abs{P}^2 + \abs{S}^2}\right) \ = [-0.062, 0.062] \ ,
    \label{eq:Smumurange}
\end{equation}
where we take $\phi_s^{\rm NP}$ from \eqref{eq:phisNP}. The range indicates the extreme values reached for either $|S|=0$ or $|P|= 0$. A measurement outside this range would clearly indicate additional sources of CP violation which could enter as complex phases in $\cminus_{10}, \;\cminus_{P}$ and $\cminus_S$. By the time such a measurement is available, we will have a much sharper picture of $\phi_s^{\rm NP}$ from $B_s\to J/\psi \phi$ \cite{DeBruyn:2022zhw}. In addition, the semileptonic rare $B_{(s)}$ decays will allow a determination of $\cminus_{10}$. If this coupling is then found to be real, a measurement of $\smumu$ would be crucial to find possible CP-violating (pseudo)-scalar interactions. In this case, a measurement outside the (updated) range in \eqref{eq:Smumurange} would unambiguously indicate additional sources of new (pseudo)-scalar interactions.

\begin{figure}
    \centering
    \subfloat[]{\includegraphics[height=0.30\textwidth]{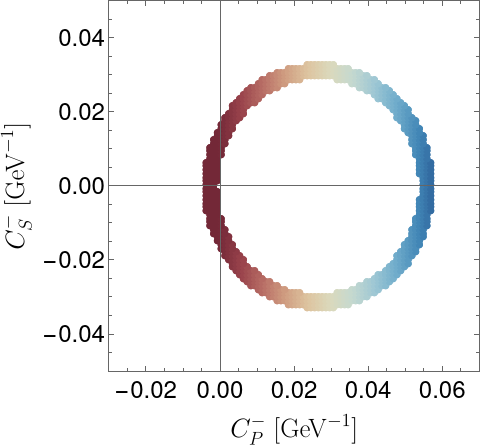}\label{fig:fig4a}}
    \hfill
    \subfloat[]{\includegraphics[height=0.30\textwidth]{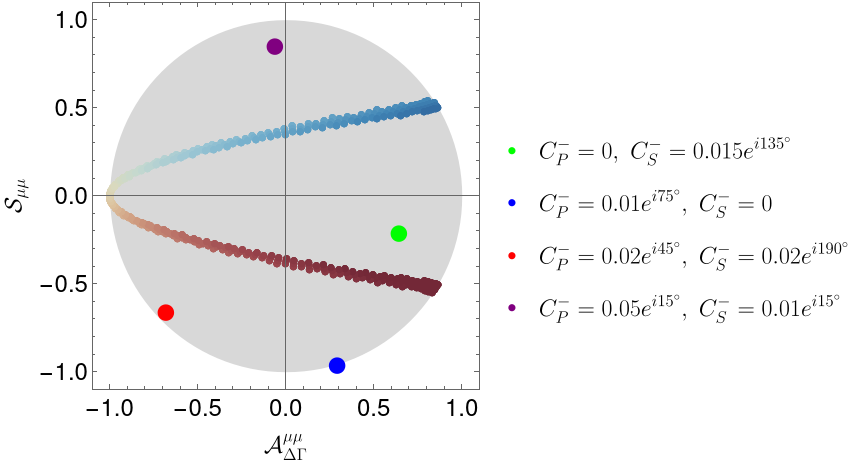}\label{fig:fig4b}}
    \caption{Correlations for the complex $\cminus_{10}$ benchmark scenario \eqref{eq:c10minus_benchmark}. (a) Constraints on real $\cminus_P$ and $\cminus_S$ from $\overline{\mathcal{B}}(B_s \to \mu^+\mu^-)$, and (b) the corresponding correlation between $\adg$ and $\smumu$ indicated with the same colour coding. The coloured points in (b) show the correlation for complex $\cminus_P$ and/or $\cminus_S$ which spans the whole unit disk as discussed in the text.}
    \label{fig:corrplotADGSmumu}
\end{figure}

In the more exciting case that new CP-violating $\cminus_{10}$ couplings are found, it would be important to explore whether such effects are also present in the (pseudo)-scalar interactions. We stress that this cannot simply be established by a global fit for two main reasons. First, the semileptonic decays are insensitive to (pseudo)-scalar couplings and second because $B_s\to \mu^+\mu^-$ does not offer enough observables to pin down new (pseudo)-scalar interactions \cite{Fleischer:2017yox}. Therefore, it is important to consider the $B_s\to \mu^+\mu^-$ sector separately from the semileptonic decays, as proposed in our new strategy. In order to establish new (pseudo)-scalar interactions, an analysis of $\smumu$ and $\adg$ in $B_s \to \mu^+\mu^-$ will be crucial. As discussed in the previous subsection, in this case $\overline{\mathcal{B}}(B_s \to \mu^+\mu^-)$ constrains a circle in the $\cminus_P$--$\cminus_S$ plane. Figure~\ref{fig:corrplotADGSmumu}\subref{fig:fig4a} shows this circle for our complex benchmark value of $\mathcal{C}_{10}^{-\rm NP}$ in \eqref{eq:c10minus_benchmark}. This constraint can be translated into the $\adg$--$\smumu$ plane, yielding the red and blue C-shaped region in Fig.~\ref{fig:corrplotADGSmumu}\subref{fig:fig4b}. 
The region shows which values of $\adg$ and $\smumu$ are consistent with data on $\overline{\mathcal{B}}(B_s \to \mu^+\mu^-)$ when $\cminus_P$ and $\cminus_S$ are real but $\cminus_{10}$ is complex. The resulting strong correlation between $\adg$ and $\smumu$ provides a narrow target region for the search for new CP-violating sources that enter through the (pseudo)-scalar Wilson coefficients.

Moreover, if $\cminus_P$ and/or $\cminus_S$ have complex phases, $\adg$ and $\smumu$ are no longer restricted to this region but can take any values within the unit disk. If future measurements were to fall outside the target region, we would have evidence for a complex phase in $\cminus_P$ and/or $\cminus_S$. In order to illustrate this exciting feature, we have marked four coloured points in Fig.~\ref{fig:corrplotADGSmumu}\subref{fig:fig4b} for several complex coefficients $\cminus_P$ and/or $\cminus_S$.

\begin{figure}
    \centering
    \subfloat[${\mathcal{C}_{10}^{- \rm NP} = 0.30 |C_{10}^{\rm SM}| e^{i \: 60^\circ}}$]{\includegraphics[height=0.26\textwidth]{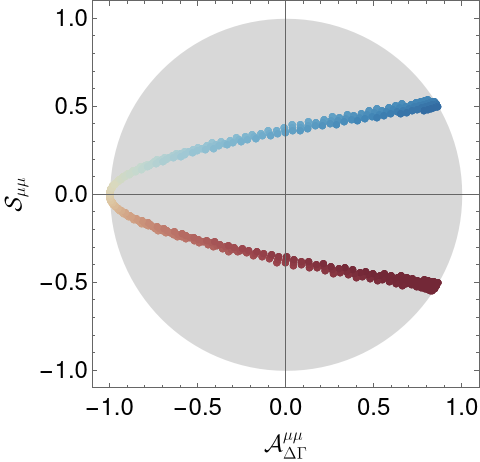}}
    \hfill
    \subfloat[${\mathcal{C}_{10}^{- \rm NP} = 0.15 |C_{10}^{\rm SM}| e^{-i \: 60^\circ}}$]{\includegraphics[height=0.26\textwidth]{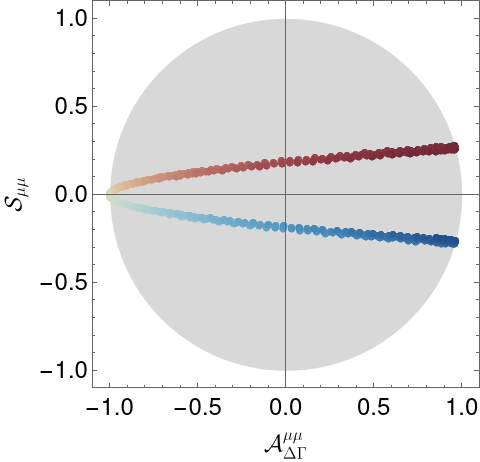}}
    \hfill
    \subfloat[${\mathcal{C}_{10}^{- \rm NP} = 0.45 |C_{10}^{\rm SM}| e^{i \: 90^\circ}}$]{\includegraphics[height=0.26\textwidth]{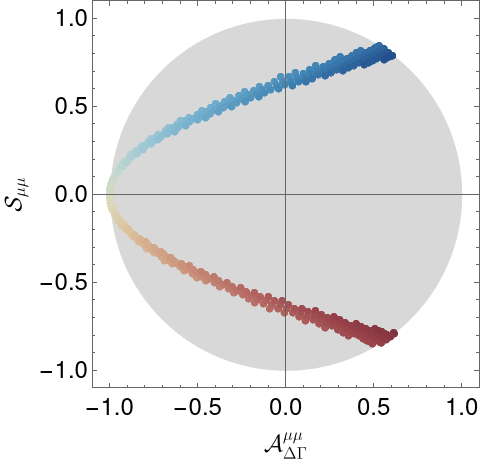}}
    \caption{Correlations between $\adg$ and $\smumu$ for different values of $\cminus_{10}$.}
    \label{fig:realCPandCSfingerprint_moreplots}
\end{figure}

We conclude that through the surprisingly strong correlation between $\adg$ and $\smumu$, we may reveal possible CP-violating effects in the (pseudo)-scalar sector.
We stress that our method is robust with respect to the value of $\cminus_{10}$ extracted from the semileptonic $B$ decays. In Fig.~\ref{fig:realCPandCSfingerprint_moreplots}, we show the target regions that result from three different benchmark scenarios for $\mathcal{C}_{10}^{- \rm NP}$, which show the power of this approach: after fixing $\cminus_{10}$, the branching ratio $\overline{\mathcal{B}}(B_s \to \mu^+\mu^-)$ constrains a very specific region in the $\adg$--$\smumu$ plane. Without fixing $\cminus_{10}$, the whole unit disk is allowed. 

\begin{figure}
    \centering
    \subfloat[$\mathcal{C}_{10}^{- \rm NP} = 0.30 |C_{10}^{\rm SM}| e^{i \: 60^\circ}$]{\includegraphics[width=0.29\textwidth]{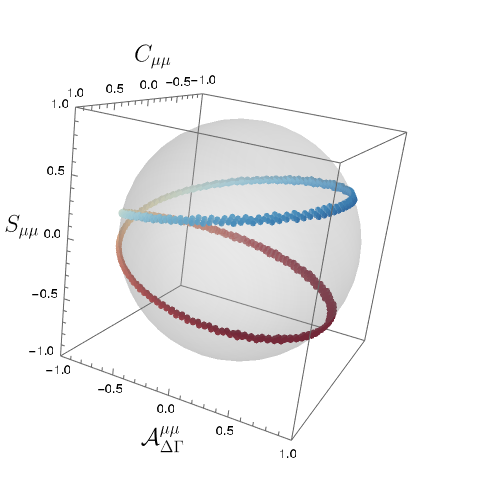}}
    \hfill
    \subfloat[$\mathcal{C}_{10}^{- \rm NP} = 0.15 |C_{10}^{\rm SM}| e^{-i \: 60^\circ}$]{\includegraphics[width=0.29\textwidth]{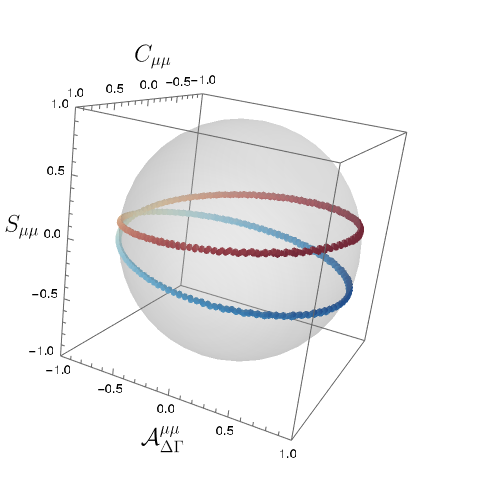}}
    \hfill
    \subfloat[$\mathcal{C}_{10}^{- \rm NP} = 0.45 |C_{10}^{\rm SM}| e^{i \: 90^\circ}$]{\includegraphics[width=0.29\textwidth]{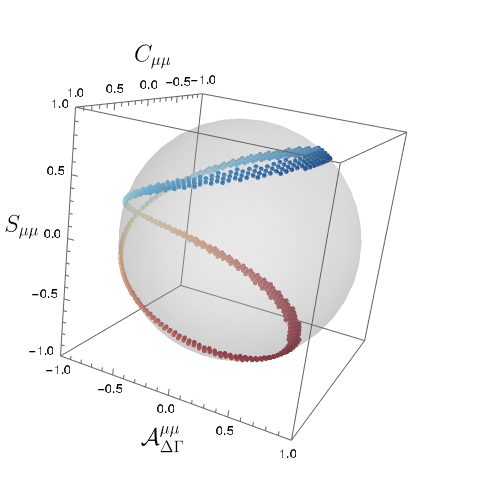}}
    \caption{3D correlations between $\adg$, $\smumu$, and $\cmumu$ for different values of $\cminus_{10}$.}
    \label{fig:3Dplot}
\end{figure}

For completeness, we also consider the correlations with $\cmumu$, defined in \eqref{eq:Cmumu}. This observable, which is very challenging to measure, is only non-zero if scalar couplings are present. In Fig.~\ref{fig:3Dplot}, we show the unit spheres in the $\adg$--$\smumu$--$\cmumu$ space for three different values of $\cminus_{10}$ given by \eqref{eq:sphere}. Again, the coloured contours are those allowed for real (pseudo)-scalar couplings coefficients. The projection along the $\cmumu$ axes corresponds to the 2D plots presented above.

\begin{figure}[t]
\centering
\includegraphics[width=0.93\textwidth]{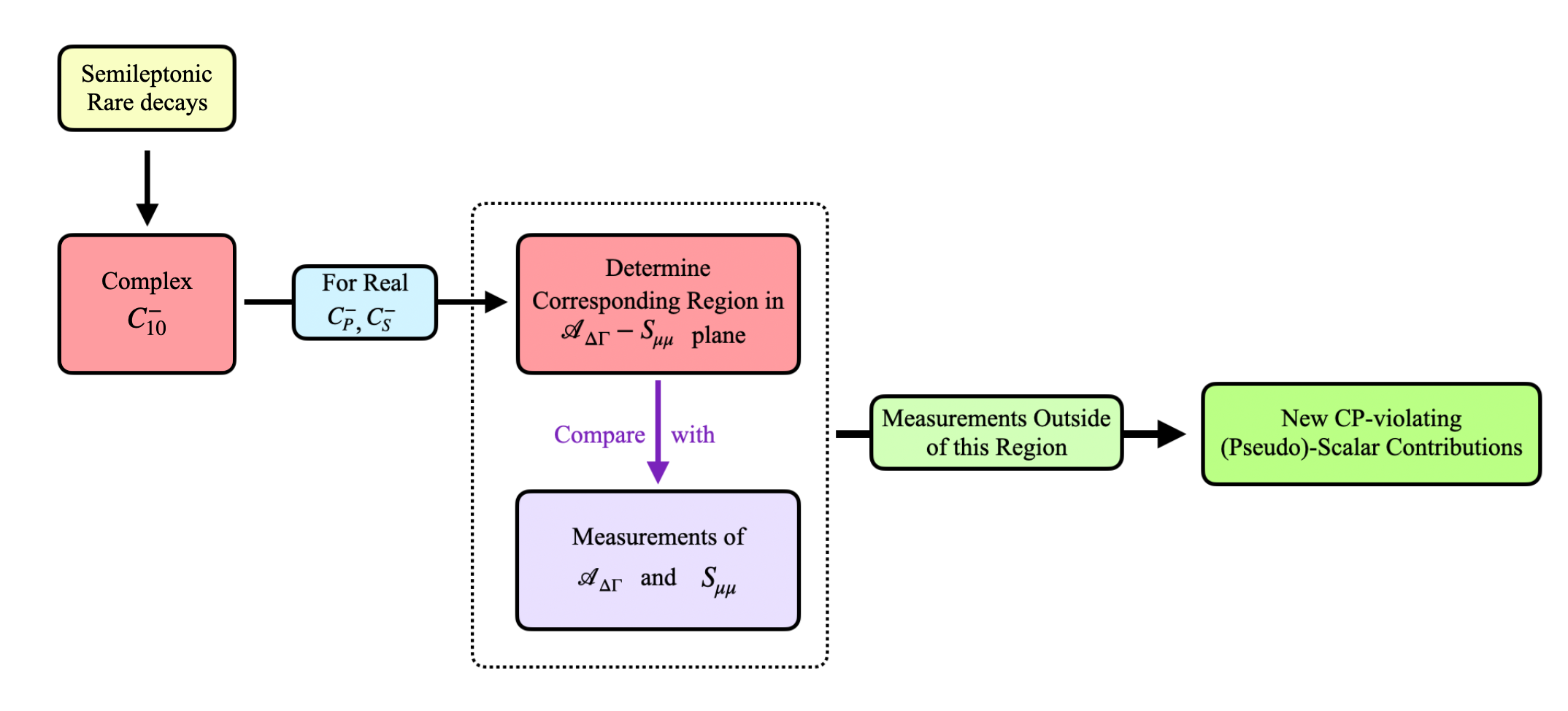}
    \caption{Illustration of our strategy to detect the presence of CP-violating phases in $\cminus_P$ and/or $\cminus_S$.}
    \label{fig:flow}
\end{figure}

In summary, the strong correlation between $\adg$ and $\smumu$, in combination with the branching ratio $\overline{\mathcal{B}}(B_s \to \mu^+\mu^-$), allows us to detect the presence of possible CP-violating phases in $\cminus_P$ and/or $\cminus_S$. We illustrate the procedure in the flowchart of Fig.~\ref{fig:flow}:  Starting from the rare semileptonic $B\to K$ and $B_{(s)} \to K^*(\phi)\mu^+\mu^-$ decays, in the future, the complex $\cminus_{10}$ can be extracted from branching ratios, angular observables and CP asymmetry measurements. This will allow us to constrain a region in the $\adg$--$\smumu$ plane. Once future time-dependent $B_s\to \mu^+\mu^-$ data are available, as a final step, we can compare our target region with data to determine if there are as well new CP-violating contributions in the $\cminus_S$ and/or $\cminus_P$ coefficients.

\section{Conclusions and Outlook}
\label{ch:conclusions}

We have presented a new strategy to determine whether there are new CP-violating (pseudo)-scalar interactions by fully exploiting the rare $B_s\to \mu^+\mu^-$ decay. We find that especially future measurements of the $\adg$ and $\smumu$ observables have great potential in revealing New Physics. 

In our method, we use $\cminus_{10}$ as an external input. We have demonstrated that the rare semileptonic $B_{(s)}\to K\mu^+\mu^-$ and $B\to K^* (\phi)\mu^+\mu^-$  decays are insensitive to new (pseudo)-scalar coefficients, which allows the extraction of the complex $\cminus_{10}$ independently of $\cminus_P$ and $\cminus_S$. Using several benchmark scenarios, we then find surprisingly constrained target regions in the $\adg$--$\smumu$ plane assuming NP entering in $\cminus_{P}$ and $\cminus_{S}$ to be CP-conserving. Measurements outside these regions would indicate new CP-violating phases in the (pseudo)-scalar coefficients. This method provides an efficient way to identify possible new sources of CP violation in the (pseudo)-scalar sector.

Given the potential of CP-violating observables in $B_s \to \mu^+\mu^-$ to reveal NP, we strongly encourage the experimental community to perform the corresponding measurements. Pioneering, untagged time-dependent analyses of the effective lifetime by LHCb, ATLAS, and CMS are already available, leading to first determinations of $\adg$. As a next step, it would be crucial to perform also tagged time-dependent analyses, allowing the extraction of the observable $\smumu$. 

By the time these challenging measurements are available, we will have a clearer picture of possible NP effects. In particular, $\phi_s^{\rm NP}$ will have been constrained with much higher precision, which may even establish a non-zero value of that phase. Moreover, measurements in the semileptonic rare $B$ decays may already have brought us into the exciting scenario where NP effects are revealed in the coefficient $\cminus_{10}$. Either way, the observables in $B_s\to \mu^+\mu^-$ will play a key role in probing new CP-violating effects in (pseudo)-scalar interactions and determining the full dynamics of this fascinating rare decay.

\section*{Acknowledgements}
This research has been supported by the Netherlands Organisation for Scientific Research (NWO).

\bibliography{refs.bib}
\end{document}